\documentclass[12pt]{article}
\usepackage{amssymb}
\usepackage{epsfig}

\catcode`\@=11
\def\numberbysection{\@addtoreset{equation}{section}
        \def\theequation{\thesection.\arabic{equation}}}

\def\beq{\begin{equation}}
\def\eeq{\end{equation}}
\numberbysection
\begin{document}
\begin{titlepage}
\begin{center}
\hfill DFF  1/12/06 \\
\vskip 1.in {\Large \bf Entanglement entropy and 2d black holes }
\vskip 0.5in P. Valtancoli
\\[.2in]
{\em Dipartimento di Fisica, Polo Scientifico Universit\'a di Firenze \\
and INFN, Sezione di Firenze (Italy)\\
Via G. Sansone 1, 50019 Sesto Fiorentino, Italy}
\end{center}
\vskip .5in
\begin{abstract}
We explore the method of entanglement entropy applied to $2d$ black
holes. We introduce a solvable model of a real scalar field with
finite volume and lattice spacing in terms of $N$ coupled mechanical
oscillators and compute its entanglement entropy in many cases. The
large $N$ limit of this scheme, with finite lattice spacing, should
reproduce the Bekenstein-Hawking formula for black hole entropy.
\end{abstract}
\medskip
\end{titlepage}
\pagenumbering{arabic}
\section{Introduction}

Recently, remarkable progresses have been made in order to
understand the microscopic origin of the thermodynamic behavior of
black holes \cite{1}-\cite{2} ( for a review see \cite{a1} ). Let us
mention the rigorous proof of black hole entropy achieved in
superstring theory, using the D-brane technology \cite{3}-\cite{6}.
Analogously in loop quantum gravity \cite{7} black hole entropy has
been identified with the logarithm of the total number of all
different spin network states for a fixed eigenvalue of the area
operator \cite{8}.

The derivations made in these candidates for quantum gravity
strongly depend on the details of these theories. However it is
generally believed that the proportionality of the black hole
entropy with the horizon area is more universal and not dependent on
the details of the theory.

In this direction we can recall 't Hooft  \cite{9} identification of
black hole entropy with the statistical entropy of a thermal gas of
quantum particles in the Hartle-Hawking state \cite{10}-\cite{11}
with Dirichlet boundary conditions just outside the horizon ( the so
called brick wall model ). He found, in the WKB approximation, apart
from the expected volume dependent thermodynamic quantities
describing hot fields in an almost flat space, additional surface
contributions proportional to the horizon area. These extra
contributions are also proportional to $\alpha^{-2}$, where $\alpha$
is the proper altitude of the wall above the gravitational radius
and they diverge in the $\alpha \rightarrow 0$ limit. These
divergencies have been recognized as quantum corrections to the
Bekenstein-Hawking formula that can be reabsorbed in the
renormalization of the one loop effective gravitational lagrangian.
For a specific choice of $\alpha$, which is generally at the Planck
scale, 't Hooft has been able to reproduce the Bekenstein-Hawking
formula with the correct coefficient.

The original formulation of the brick wall model has been corrected
in \cite{12}-\cite{15}. Recently it has been reconsidered in the
special case of ($1+1$) dimensions \cite{16}, because it can be
solved exactly without any approximation.

A concurrent method has been developed in \cite{17}-\cite{19}, i.e.
the entanglement method, which we are going to reconsider in this
article always in the solvable ($1+1$) dimensional case. This method
is a strong candidate for a model independent derivation of the
origin of black hole entropy, since it defines a statistical entropy
measuring the loss of information induced by a spatial division of a
system.

This division distinguishes between the inaccessible quantum degrees
of freedom made unphysical by the presence of black hole and the
residual physical ones. It is expected that, independently from the
details of the theory, entanglement entropy is proportional to the
area of the boundary of the spatial division of the system.

The advantage of ($1+1$) dimensions is that there is a coordinate
transformation making the background metric conformally flat, and
therefore the quantization of a scalar field in the background of a
black hole can be reduced to the free case.

We point out that the natural spatial division which we adopt to
compute black hole entropy is not separating the interior of black
hole from the external region, but, since the coordinate
transformation flattening the metric carries an extra length scale
$\alpha$, the wall proper altitude, is distinguishing between the
interior of the brick wall and the external region to the wall.

The quantization of a free scalar field is solved by introducing the
proper modes, which are definitely non local with respect to the
spatial division of the scalar field. Therefore the vacuum wave
function carries correlations between the inaccessible degrees of
freedom inside the brick wall and the residual physical ones. Such a
correlation is the mechanism responsible for the production of a non
trivial entropy, which is similar to the brick wall entropy
calculated by 't Hooft.

To simplify the calculation it is necessary quantizing the scalar
field reduced to a chain of $N$ coupled oscillators with periodic
boundary conditions ( finite volume and finite lattice spacing ),
i.e. a system with a finite number $N$ of degrees of freedom. We are
then able to carry out the entropy calculation in many cases of
entanglement applying the general method of ref. \cite{17}. In all
these cases the entropy is finite and dependent only on the
frequencies of the normal modes.

However only in the $N \rightarrow \infty$ limit and finite lattice
spacing it is possible to determine a surface entropy contribution (
proportional to the area of the horizon  and diverging as $a^{-2}$,
with $a$ the lattice spacing, in four dimensions ). In this limit we
have to solve exactly the spectrum of an integral operator, but we
haven't found any better insight with respect to the qualitative
features depicted in \cite{17}-\cite{18}, so we leave its solution
to a future study.

\section{The model}

The brick wall model \cite{9} has shown that black hole entropy can
be obtained in an heuristic way from the quantization of a scalar
field on curved background extended to a finite temperature state,
in which certain quantum degrees of freedom are lost because of the
presence of black hole. The identification of these degrees of
freedom can be realized with the entanglement method that we are
going to apply to the simplest case of ($1+1$) dimensions.

In this article we limit ourself to consider a scalar field immersed
in a background metric $\{ M, g \}$ in $1+1$ dimensions. The action

\beq S = \int d^{1+1} x \sqrt{ |g| } \left( \frac{1}{2} g^{\mu\nu}
\partial_\mu \phi \partial_\nu \phi - \frac{1}{2} m^2 \phi^2 \right)
\label{21} \eeq

implies the Klein-Gordon equation on a curved space

\beq \frac{1}{\sqrt{|g|}} \partial_\mu ( g^{\mu\nu} \sqrt{|g|}
\partial_\nu ) \phi + m^2 \phi = 0 \label{22} \eeq

From the action $S$ (\ref{21}) we can derive the corresponding
energy-momentum tensor

\beq \theta_{\mu\nu} (x) = \frac{2}{\sqrt{|g|}} \frac{
\partial S }{ g^{\mu\nu} (x) } = - \phi
\partial_\mu \partial_\nu \phi + \frac{1}{2} g_{\mu\nu} \left(
g^{\rho\sigma} \phi \partial_\rho
\partial_\sigma \phi + m^2 \phi^2 \right) \label{23} \eeq

We restrict the choice of the background metric to the case of a
classical ($1+1$) dimensional black hole

\beq ds^2 = f(r) dt^2 - \frac{1}{f(r)} dr^2 \label{24} \eeq

$f(r)$ is an arbitrary function constrained to the following
properties:

\begin{eqnarray}
& \ &  \lim_{r\rightarrow \infty} f(r) = 1  \nonumber \\
& \ & f(r) { \rm \ has \ a \ simple \ zero \ at \ r = r_0 }
\label{25}
\end{eqnarray}

where $r_0$ is the horizon with positive surface gravity

\beq k_0 = \frac{1}{2} f'(r_0) > 0 \label{26} \eeq

The classical equation of motion for the scalar field is

\beq \frac{1}{f(r)}
\partial^2_t \phi - \partial_r [ f(r) \partial_r \phi ] + m^2 \phi^2
= 0 \label{27} \eeq

In the coordinates ($t,y$) with "tortoise" coordinate $y$ defined by

\beq y = y(r) \ \ \ \ \frac{dy}{dr} = \frac{1}{f(r)} \label{28} \eeq

the black hole metric becomes conformally flat. Notice that
$f(r_0)=0$ and $f'(r_0) = 2 k_0 >0$ implies that for $r\rightarrow
r_0$ at the leading order

\beq y(r) \sim \frac{1}{2 k_0} ln 2 k_0 ( r- r_0 ) \ \ \ \ \ f(r)
\sim e^{2 k_0 y(r)} \approx 2 k_0 ( r - r_0 ) \label{29} \eeq

and therefore, containing a logarithm, it introduces an auxiliary
scale of length $1/2 k_0$. Moreover the coordinate transformation is
singular near the horizon, hence in the $y$ coordinates it doesn't
make sense introducing a spatial division of the system coinciding
with black hole horizon.

Let $r_1 > r_0$ be the radius of our alternative spatial division of
the system, analogous to 't Hooft brick wall, and let us calculate
the proper altitude of the wall $\alpha$ over the horizon depending
on $ \delta r = r_1- r_0 $:

\beq \alpha = \int^{r_1}_{r_0} f^{-\frac{1}{2}} dr \ \ \ \ \delta r
= r_1 - r_0 = \frac{1}{2} k_0 \alpha^2 \label{210} \eeq

In the coordinates $(t,y)$ the Klein-Gordon equation (\ref{27})
reduces to its flat form in the massless limit ( $m=0$ )

\beq ( \partial^2_t - \partial^2_y ) \phi = 0 \label{211} \eeq

and we have the problem of quantizing a free scalar field.

This length scale $\alpha$ is also present in the brick wall model,
where it can be proved to be of the Planck scale order and it allows
to distinguish the regions

\begin{eqnarray}
& \ & y < y(r_1) \ \ \ \ \ \ \  \ r_0 < r < r_1 \nonumber \\
& \ & y > y(r_1) \ \ \ \ \ \ \ r > r_1 \label{212} \end{eqnarray}

It is aim of our article to hypothesize that the degrees of freedom
freezed by the presence of black hole are situated in the interior
of the brick wall for $y<y(r_1)$, i.e. they are superficial, while
the residual physical ones are always external to the brick wall for
$y >y(r_1)$. In this sense we distinguish ourself from the
literature that usually identifies them in the interior of black
holes.

We notice that in the $y$ coordinates in two dimensions this spatial
division between frozen and physical degrees of freedom seems the
only possible choice. In greater dimensions or in the massive case
it is not possible to avoid the presence of a background field and
the analysis appears to be more intricate.

\section{Scalar field as a finite chain}

Before quantizing the scalar field we must restrict ourself for
simplicity to a model of it in which both the volume and the lattice
spacing are finite. It is well known that the scalar field in
($1+1$) dimensions, defined on a lattice and with a finite volume,
reduces to a system of $N$ coupled mechanical oscillators.

Consider the hamiltonian

\beq H = \frac{1}{2} \sum^{N}_{A=1} [ \dot{u}^{2}_A +
\frac{v^2}{a^2} { ( u_A - u_{A+1} ) }^2 + \omega^2_0 u^2_A ]
\label{} \label{31} \eeq

where the single oscillator variables $u_A, A = 1,..,N$ are real and
we impose periodic boundary conditions $ ( N+1 \equiv 1 ) $.

Such an hamiltonian can be diagonalized introducing the so-called
normal coordinates through a discrete Fourier transform

\begin{eqnarray}
u_A & = & \sum^{N}_{k=1} \frac{ e^{i\tilde{k} ( k ) a_A }}{\sqrt{N}}
q_k \ \ \ \ \ \
\tilde{k} = \frac{2\pi}{N a} k \ \ \ a_A = a \cdot A \nonumber \\
\dot{u}_A & = & \sum^{N}_{k=1} \frac{ e^{-i\tilde{k} ( k ) a_A
}}{\sqrt{N}} p_k \label{32}
\end{eqnarray}

or inversely

\begin{eqnarray}
q_k & = & \sum^{N}_{A=1} \frac{ e^{-i\tilde{k} ( k ) a_A
}}{\sqrt{N}} u_A \ \ \ \ \ \
\nonumber \\
p_k & = & \sum^{N}_{A=1} \frac{ e^{i\tilde{k} ( k ) a_A }}{\sqrt{N}}
\dot{u}_A \label{33}
\end{eqnarray}

The inversion is obtained by recalling the discrete completeness
relation

\beq \sum_{n=1}^{N} \frac{ e^{i n ( \tilde{k} - \tilde{k'}) a }}{N}
= \delta_{ \tilde{k},\tilde{k'} } \label{34}\eeq

The reality condition for $u_A, A= 1,...N$ is assured by the
following necessary condition

\begin{eqnarray}
q_k^{*} & = & q_{-k} \nonumber \\
p_k^{*} & = & p_{-k} \label{35}
\end{eqnarray}

By introducing the parameterization $( q_k, p_k )$ the Hamiltonian
(\ref{31}) can be recast in the following diagonal form

\beq H = \frac{1}{2} \sum_k ( {| p_k |}^2 + \omega^2_k {| q_k |}^2 )
\label{36} \eeq

where $ \omega_k $ is given by the dispersion relation

\beq \omega_k^2 = \omega^2_0 + 4\frac{v^2}{a^2} sin^2
\frac{\tilde{k}(k) a}{2} \eeq

The information about the original coupling between the oscillators
is now contained in the dispersion relation. The massless limit is
reached by putting $\omega_0 = 0$.

Such a diagonal form of the Hamiltonian is ready for the
quantization procedure of the coupled system of $N$ oscillators.

The canonical quantization rules for the variables $u_A$

\begin{eqnarray}
& \ & [ u_A , \dot{u}_B ] = i \hbar \delta_{AB} \nonumber \\
& \ & [ u_A, u_B ] = [ \dot{u}_A , \dot{u}_B ] = 0 \label{38}
\end{eqnarray}

can be extended to the normal coordinates $q_k$, since they are
obtained with their conjugate momenta $p_k$ from $u_A, \dot{u}_A$
through an unitary transformation (\ref{33})

\begin{eqnarray}
& \ & [ q_k , p_{k'} ] = i \hbar \delta_{k k'} \nonumber \\
& \ & [ q_k , q_{k'} ] = [ p_k , p_{k'} ] = 0 \label{39}
\end{eqnarray}

Now $u_A$ and $\dot{u}_A$ are promoted to hermitian operators,
$u^{\dagger}_A = u_A$, and the corresponding normal coordinates have
the following hermicity properties

\begin{eqnarray}
& \ & q^{\dagger}_k = q_{-k} \nonumber \\
& \ & p^{\dagger}_k = p_{-k} \label{310}
\end{eqnarray}

The corresponding Hamiltonian operator, written in terms of the
normal coordinates, takes the following form

\beq H = \frac{1}{2} \sum_k ( p_k p_k^{\dagger} + \omega^2_k q_k
q_k^{\dagger} ) \label{311} \eeq

The quantization procedure is simplified if we introduce the
creation and destruction operators $a^{\dagger}_k$ and $a_k$ :

\begin{eqnarray}
a_k & = & \frac{1}{\sqrt{2 \hbar \omega_k }} ( \omega_k q_k + i
p^{\dagger}_k ) \nonumber \\
a^{\dagger}_k & = & \frac{1}{\sqrt{2 \hbar \omega_k }} ( \omega_k
q^{\dagger}_k - i p_k ) \label{312}
\end{eqnarray}

or inversely

\begin{eqnarray}
q_k & = & \frac{\hbar}{\sqrt{2 \omega_k }} ( a_k +
a^{\dagger}_{-k} ) \nonumber \\
p_k & = & i \frac{\hbar}{\sqrt{2 \omega_k }} ( a_{-k} -
a^{\dagger}_{k} ) \label{313}
\end{eqnarray}

From the definition of $a_k, a^{\dagger}_k$ we immediately get

\begin{eqnarray}
& \ & [ a_k , a^{\dagger}_{k'} ] = \delta_{k k'} \nonumber \\
& \ & [ a_k , a_{k'} ] = [ a^{\dagger}_k , a^{\dagger}_{k'} ] = 0
\label{314}
\end{eqnarray}

Since $a_{-k} \neq a^{\dagger}_k$ we can state that there are $2N$
independent operators $a_k, a^{\dagger}_k$.

We can now express $H$ in terms of the operators $a_k,
a^{\dagger}_k$ obtaining

\beq H = \sum_k \hbar \omega_k ( a^\dagger_k a_k + \frac{1}{2} )
\label{315} \eeq

The state of lowest energy $|0>$ is determined by the condition

\beq a_k | 0 > = 0 \ \ \ \ \ \  \forall k \label{316} \eeq

In the Schr$\ddot{o}$dinger representation in which the $q_k$ are
basic variables, this definition can be rewritten as a differential
equation for the variables $q_k$, solved by

\beq f( q_k ) = f_0 e^{- \frac{1}{2} \sum_k \ \omega_k q_k q_k^{*} }
\label{317} \eeq

Such an eigenstate has eigenvalue $ E_0 = \sum_k \frac{\hbar
\omega_k}{2} $. The generic eigenvalues of $H$ are given by

\beq E = E_0 + \sum_k \hbar \omega_k n_k \label{318} \eeq

Introducing the normal modes we have passed from a description in
terms of single oscillators $u_A$ to a description of collective
coordinates $q_k$.

\section{Entanglement entropy}

The quantization procedure is naturally related to the collective
coordinates, while the spatial division of the system induced by the
presence of the black hole is better described in terms of the
single oscillators $u_A$. Hence when we integrate on the single
oscillator degrees of freedom contained inside the wall for
$y<y(r_1)$, clarifying where is the loss of degrees of freedom
induced by the black hole, there is a non trivial coupling given by
the vacuum wave function between frozen and physical degrees of
freedom. This correlation is the ultimate reason for the presence of
a non trivial entropy.

The vacuum wave function

\beq \psi_0 ( u_A ) = f_0 e^{- \frac{1}{2} \sum_k \omega_k q_k
q_k^{*} } \label{41} \eeq

rewritten in the single oscillator coordinates $u_A$ (\ref{32})
becomes

\begin{eqnarray} q_k & = & \sum_{A=1}^N \frac{e^{-i \tilde{k} (k) a_A}}{ \sqrt{N} }
u_A \nonumber \\
\psi_0 ( u_A ) & = & f_0 \ e^{ - \frac{1}{2} M_{AB} u^A u^B }
\label{42}
\end{eqnarray}

where

\beq M_{AB} = \frac{1}{2N} \omega_k e^{i \tilde{k} (k) ( a_B - a_A
)} + ( A \leftrightarrow B ) \label{43} \eeq

Imposing the normalization condition of the wave function we get

\beq \psi_0 ( u_A ) = {\left[ det \frac{M}{\pi}
\right]}^{\frac{1}{4}} e^{ - \frac{1}{2} M_{AB} u^A u^B } \label{44}
\eeq

$M_{AB}$ is a positive definite quadratic form; this property will
be crucial in the following.

The density matrix for the vacuum is given by

\beq \rho = | \psi_0 >< \psi_0 | \label{45} \eeq

and in the Schr$\ddot{o}$dinger representation it becomes

\beq \rho \{ u^A, u'^B \} = {\left[ det \frac{M}{\pi}
\right]}^{\frac{1}{2}} e^{ - \frac{1}{2} M_{AB} ( u^A u^B + u'^A
u'^B ) } \label{46}\eeq

Now we take into account the presence of the black hole. We have
said that, according to the hypothesis of section $2$, in the region
contained between the horizon and the brick wall, the quantum
information is inaccessible. We can therefore obtain a reduced
density matrix $ \rho_{red} $ for the degrees of freedom outside the
brick wall, integrating the single oscillator coordinates $u_A$ that
are situated in the region $y < y(r_1) $:

\begin{eqnarray}
& \ & \rho_{red} ( u^a, u'^b ) = \int \prod_{\alpha} du^{\alpha}
\rho ( u^a, u^\alpha, u'^b, u^\alpha ) \nonumber \\
& \ & = {\left[ det \frac{M}{\pi} \right]}^{\frac{1}{2}} e^{ -
\frac{1}{2} M_{ab} ( u^a u^b + u'^a u'^b ) } \int \prod_{\alpha}
du^{\alpha} e^{ [ - M_{\beta\gamma} u^\beta u^\gamma - M_{\alpha a}
( u^a + u'^a ) u^\alpha ]} \label{47} \end{eqnarray}

where the matrix $M_{AB}$ is decomposed as

\beq M_{AB} = \left[ \begin{array}{cc} M_{ab} & M_{a\alpha} \\
M_{\alpha b} & M_{\alpha\beta} \end{array} \right] \label{48} \eeq

and the inverse matrix $M^{AB}\ ( M^{AB} M_{BC} = \delta^A_C ) $

\beq M^{AB} = \left[ \begin{array}{cc} M^{ab} & M^{a\alpha} \\
M^{\alpha b} & M^{\alpha\beta} \end{array} \right] \label{49} \eeq

If we introduce as in ref. \cite{17} the following extra notations,
i.e. $ \widetilde{M}^{ab} $ as inverse of $ M_{ab} $ , $
\widetilde{M}_{ab} $ as inverse of $ M^{ab} $, $
\widetilde{M}^{\alpha\beta} $ as inverse of $ M_{\alpha\beta} $ e $
\widetilde{M}_{\alpha\beta} $ as inverse of $ M^{\alpha\beta} $,
performing the quadratic integration we end up with the following
formula

\begin{eqnarray} \rho_{red} ( u^a, u'^b ) = { \left[ det \left(
\frac{\widetilde{M}_{ab}}{\pi} \right) \right] }^{\frac{1}{2}} e^{
-\frac{1}{2} M_{ab} ( u^a u^b + u'^a u'^b )  } e^{ \frac{1}{4}
\widetilde{M}^{\alpha\beta} M_{\alpha a} M_{\beta b} {( u+u' )}^a {(
u+u' )}^b  } \label{410}
\end{eqnarray}

where

\beq det M_{AB} = det \widetilde{M}_{ab} \ det M_{\alpha\beta}
\label{411} \eeq

The integration is non trivial only because of the coupling given by
$M_{\alpha a}$ between the degrees of freedom inside the region $y <
y(r_1)$ and those external to it ($y>y(r_1)$).

Using the identity

\beq \widetilde{M}_{ab} = M_{ab} - M_{a \alpha}
\widetilde{M}^{\alpha\beta} M_{\beta b} \label{412} \eeq

and defining

\beq K_{ab} = \widetilde{M}_{ab} \ \ \ \ \ H_{ab} = M_{a \alpha}
\widetilde{M}^{\alpha\beta} M_{\beta b} \label{413} \eeq

the resulting density matrix is

\begin{eqnarray} & \ &  \rho_{red} \{ u^a, u'^b \} = { \left[ det \left(
\frac{K_{ab}}{\pi} \right) \right] }^{\frac{1}{2}} e^{ - \frac{1}{2}
K_{ab} ( u^a u^b + u'^a u'^b )  } \nonumber \\
& \ & e^{  - \frac{1}{4} H_{ab} {( u - u' )}^a {(u-u')}^b  }
\label{414} \end{eqnarray}

In the special case of two degrees of freedom, of which one inside
and the other outside the wall, the density matrix (\ref{414})
becomes

\beq \rho_{red} ( u, u' ) = { \left[ \frac{K}{\pi} \right]
}^{\frac{1}{2}} e^{ - \frac{1}{2} K ( x^2 + x'^2 ) - \frac{1}{4} H (
x - x' )^2 } \label{415} \eeq

The entropy of this mixed state depends only on the ratio

\beq \lambda = \frac{H}{K} \label{416} \eeq

and on the accessory variable

\beq \mu = 1 + \frac{2}{\lambda} - 2 {\left[ \frac{1}{\lambda}
\left( 1 + \frac{1}{\lambda} \right) \right]}^{\frac{1}{2}}
\label{417} \eeq

through the formula

\beq S = - \frac{ \mu ln \mu + ( 1 - \mu ) ln ( 1 - \mu ) }{ 1 - \mu
} \label{418} \eeq

In the general case it is possible to extrapolate the general
formula for the entropy by bringing the density matrix (\ref{414})
to the following diagonal form

\beq \rho_{red} ( u^a, u'^b ) = \prod_a \pi^{-\frac{1}{2}} e^{ [ -
\frac{1}{2} ( u^a u^a + u'^a u'^a ) - \frac{1}{4} \lambda_a ( u-u'
)^a ( u-u' )^a ] } \label{419} \eeq

where the $\lambda_a$ are the eigenvalues of $H_{ab}$, in the basis
where $ K_{ab} $ is the identity matrix, and in general they are
recognizable as the eigenvalues of the composition of matrices

\beq \Lambda^a_b = {( K^{-1} )}^{ac} { ( H ) }_{cb} \label{420} \eeq

This is possible since at least one of $K_{ab}$ and $H_{ab}$ is
associated to a positive definite quadratic form, with positive
eigenvalues. Hence with a series of transformations $K_{ab}$ can be
diagonalized simultaneously with the matrix $ H_{ab} $.

Finally, the entropy associated to a system of many degrees of
freedom is given by

\beq S = - \sum_a \frac{ \mu_a ln \mu_a + ( 1 - \mu_a ) ln ( 1 -
\mu_a ) }{ 1 - \mu_a } \label{421} \eeq

where $\mu_a$ is the positive solution of

\beq \lambda_a = \frac{ 4\mu_a }{{(1-\mu_a)}^2 } \label{422} \eeq

and $\lambda_a$ are the eigenvalues of the matrix $\Lambda^a_b$.

\section{Entropy computation}

The natural application of the general theory outlined in the last
section is the model of the scalar field reduced to $N$ degrees of
freedom. We will be able to compute directly without examining the
reduced density matrix the entropy of such a system in terms of the
frequencies $\omega_k$ characterizing the normal modes of the field.

In the simplest $N=2$ case with a degree of freedom inside the brick
wall and the other outside, we obtain from the general
parameterization (\ref{43})

\begin{eqnarray}
& \ &  M_{AB} = \frac{1}{2} \left( \begin{array}{cc} \omega_1 +
\omega_2 & - \omega_1 + \omega_2 \\ - \omega_1 + \omega_2 & \omega_1
+ \omega_2 \end{array} \right) \nonumber \\
& \ &  M^{AB} = \frac{1}{2\omega_1 \omega_2} \left(
\begin{array}{cc} \omega_1 + \omega_2 & \omega_1 - \omega_2 \\
\omega_1 - \omega_2 & \omega_1 + \omega_2 \end{array} \right)
\label{51}
\end{eqnarray}

and therefore the matrices $K$ and $H$ reduce to pure numbers

\beq K = \frac{2\omega_1 \omega_2}{\omega_1+\omega_2} \ \ \ \  H =
\frac{ {( \omega_1 - \omega_2 )}^2 }{ 2( \omega_1 + \omega_2 ) }
\label{52} \eeq

Finally the eigenvalue $\lambda$ (\ref{416}) from which the mixed
state entropy depends on is

\beq \lambda = \frac{1}{4} \frac{ {( \omega_1 - \omega_2 )}^2 }{
\omega_1 \omega_2 } \label{53} \eeq

We can go on in generalizing this result to the case $(1, N-1)$,
i.e. in which one degree of freedom is observable and $N-1$ are the
inaccessible ones.

Let us take for example the case $N=3$ with $(1,2)$, we obtain in
particular

\begin{eqnarray}
q_k & = & S_{kA} u_A \ \ \ \ \ \ \ \ M_{AB} = Re \ (
S^{\dagger}_{Ak} \omega_k
S_{kB} ) \nonumber \\
q^{*}_k & = & u^T_A S^{\dagger}_{Ak} \label{54} \end{eqnarray}

where $Re$ implies taking the real part and we compute the
transformation matrices $S_{kA}$ as

\beq S_{kA} = \frac{1}{\sqrt{3}} \left( \begin{array}{ccc} \alpha &
\alpha^2 & 1 \\ \alpha^2 & \alpha & 1 \\ 1 & 1 & 1
\end{array} \right) \ \ \ \ \ \ \ \ S^{\dagger}_{Ak} =
\frac{1}{\sqrt{3}} \left(
\begin{array}{ccc} \alpha^2 & \alpha & 1 \\ \alpha & \alpha^2 & 1 \\ 1 &
1 & 1 \end{array} \right) \ \ \ \ \alpha = e^{-\frac{2\pi i}{3}}
\label{55} \eeq

The specific form of $M_{AB}$ is

\beq M_{AB} = \left[ \begin{array}{c|c} M_{ab} & M_{a\alpha} \\
\hline M_{\alpha b} & M_{\alpha\beta} \end{array} \right] = \left[
\begin{array}{c|cc} \rho_1 & \rho_2 & \rho_2 \\
\hline \rho_2 & \rho_1 & \rho_2 \\ \rho_2 & \rho_2 & \rho_1
\end{array} \right] \label{56} \eeq

where

\begin{eqnarray}
& \ & \rho_1 = \frac{1}{3} ( \omega_1 + \omega_2 + \omega_3 )
\nonumber \\
& \ & \rho_2 = \frac{1}{3} \left( - \frac{(\omega_1 + \omega_2)}{2}
+ \omega_3 \right) \label{57}
\end{eqnarray}

The inverse matrix satisfies the following properties

\beq M^{AB} = \left[ \begin{array}{c|c} M^{ab} & M^{a\alpha} \\
\hline M^{\alpha b} & M^{\alpha\beta} \end{array} \right] = \left[
\begin{array}{c|cc} \delta_1 & \delta_2 & \delta_2 \\
\hline \delta_2 & \delta_1 & \delta_2 \\ \delta_2 & \delta_2 &
\delta_1 \end{array} \right] \label{58} \eeq

where

\begin{eqnarray}
& \ & \delta_1 = \frac{\rho_1+\rho_2}{(\rho_1-\rho_2)(\rho_1 +2
\rho_2)}
\nonumber \\
& \ & \delta_2 = - \frac{\rho_2}{(\rho_1-\rho_2)(\rho_1 +2 \rho_2)}
\label{59} \end{eqnarray}

In this case the matrix $K$ is still a number

\beq K = \frac{1}{\delta_1} \label{510} \eeq

while the matrix $H$ is a number given by

\beq H = M_{a\alpha} \widetilde{M}^{\alpha\beta} M_{\beta b} =
\frac{( \rho_2 \ \rho_2 )}{\rho^2_1 - \rho^2_2} \left(
\begin{array}{cc} \rho_1 & - \rho_2
\\ - \rho_{2} & \rho_1 \end{array} \right) \left(
\begin{array}{c} \rho_2 \\ \rho_2 \end{array}
\right) = \frac{ 2 \rho^2_2 }{\rho_1 + \rho_2 } \label{511} \eeq

and the eigenvalue $\lambda = \frac{H}{K}$ is given by

\beq \lambda = \frac{ 2 \rho^2_2 }{(\rho_1 - \rho_2 )(\rho_1 + 2
\rho_2) } = \frac{4}{9} \frac{ ( \omega_3 -
\frac{\omega_1+\omega_2}{2})^2}{(\omega_1+\omega_2) \omega_3}
\label{512} \eeq

It is also easy to study the case $(2,1)$:

\beq M_{AB} = \left[ \begin{array}{c|c} M_{ab} & M_{a\alpha} \\
\hline M_{\alpha b} & M_{\alpha\beta} \end{array} \right] = \left[
\begin{array}{cc|c} \rho_1 & \rho_2 & \rho_2 \\
\rho_2 & \rho_1 & \rho_2 \\ \hline \rho_2 & \rho_2 & \rho_1
\end{array} \right] \label{513} \eeq

where $\rho_1$ and $\rho_2$ are as in eq. (\ref{57}).

Analogously the inverse matrix

\beq M^{AB} = \left[ \begin{array}{c|c} M^{ab} & M^{a\alpha} \\
\hline M^{\alpha b} & M^{\alpha\beta} \end{array} \right] = \left[
\begin{array}{cc|c} \delta_1 & \delta_2 & \delta_2 \\
\delta_2 & \delta_1 & \delta_2 \\ \hline \delta_2 & \delta_2 &
\delta_1 \end{array} \right] \label{514} \eeq

In this case the matrix $K_{ab}$ becomes

\beq K_{ab} = \frac{1}{\delta^2_1 - \delta^2_2} \left(
\begin{array}{cc} \delta_1 & - \delta_2
\\ - \delta_2 & \delta_1 \end{array} \right) \label{515} \eeq

while the matrix $H_{ab}$ is

\beq H_{ab} = \frac{\rho_2^2}{\rho_1} \left(
\begin{array}{cc} 1  & 1
\\ 1 & 1 \end{array} \right) \label{516} \eeq

from which the matrix to be diagonalized  is

\beq \Lambda^a_b = {( K^{-1} )}^{ac} H_{cb} =
\frac{\rho^2_2}{\rho_1} \left(
\begin{array}{cc} \delta_1 & \delta_2
\\ \delta_2 & \delta_1 \end{array} \right)
\left(
\begin{array}{cc} 1 & 1
\\ 1 & 1 \end{array} \right) =
\frac{(\delta_1+\delta_2)\rho^2_2}{\rho_1} \left(
\begin{array}{cc} 1 & 1
\\ 1 & 1 \end{array} \right) \label{517}
\eeq

$\Lambda^a_b$ is a matrix  whose determinant is equal to zero while
the trace

\beq Tr \Lambda^a_b  = \delta_1 \rho_1 - 1 =
2\frac{\rho^2_2}{(\rho_1-\rho_2)(\rho_1+2\rho_2)}  \label{518} \eeq

from which its eigenvalues are recognizable as

\begin{eqnarray}
& \ & \lambda_1 = \frac{4}{9} \frac{ ( \omega_3 -
\frac{\omega_1+\omega_2}{2})^2}{(\omega_1+\omega_2)
\omega_3} \nonumber \\
& \ & \lambda_2 = 0 \label{519}
\end{eqnarray}

Therefore we can come to the conclusion that the cases $(1,N-1)$ and
$(N-1,1)$ are related by duality. At first sight in the case
$(N-1,1)$ it seems that we have to diagonalize a matrix of dimension
$(N-1)$, which is however singular since it contains $N-2$ null
eigenvalues and only one that coincides with the case $(1,N-1)$.

More generally we can guess the existence of a duality between the
cases $(k,N-k)$ and $(N-k,k)$, with the first case containing $k$
independent contributions to the total entropy, while the second one
contains the same eigenvalues plus extra null contributions until
completing $N-k$ eigenvalues.

\section{N = 4 case }

Let us consider the case $N=4$, that, apart from exhibiting the
usual cases of the form $ ( 1, N-1 ) $ or $ ( N-1, 1 ) $, introduces
the new non trivial case $ ( 2, 2 )$.

Let us define the matrices $ M_{AB} $ and $ M^{AB} $ as

\begin{eqnarray}
M_{AB} & = & Re ( S^{\dagger}_{Ak} A_{km} S_{mB} ) \nonumber \\
M^{AB} M_{BC} & = & \delta^A_C \label{61}
\end{eqnarray}

where the diagonal matrix $A_{km}$ has the following form

\beq A_{km} = \left( \begin{array}{cccc} \omega_1 & & & \\ &
\omega_2 & & \\ & & \omega_3 & \\ & & & \omega_4 \end{array} \right)
\label{62} \eeq

while the similitude matrix $S_{kA}$ is given by

\beq S_{kA} = \frac{1}{2} \left( \begin{array}{cccc} \alpha &
\alpha^2 & \alpha^3 & 1
 \\ \alpha^2 & 1 & \alpha^2 & 1 \\ \alpha^3 & \alpha^2 & \alpha & 1 \\ 1 & 1 & 1 & 1
 \end{array}
\right) \ \ \ \ \ \ \ S^{\dagger}_{Ak} = \frac{1}{2} \left(
\begin{array}{cccc} \alpha^3 & \alpha^2 & \alpha & 1
 \\ \alpha^2 & 1 & \alpha^2 & 1 \\ \alpha & \alpha^2 & \alpha^3 & 1 \\ 1 & 1 & 1 & 1
 \end{array}
\right) \ \ \ \ \ \ S^{\dagger} S = 1 \ \ \alpha^4 = 1
\label{63}\eeq

Computing explicitly the matrix $M_{AB}$ we obtain

\beq M_{AB} = \left( \begin{array}{cccc} \rho_1 & \rho_2 & \rho_3 & \rho_2 \\
\rho_2 & \rho_1 & \rho_2 & \rho_3 \\
\rho_3 & \rho_2 & \rho_1 &\rho_2 \\
\rho_2 & \rho_3 & \rho_2 & \rho_1 \end{array} \right) \label{64}
\eeq

with the following definitions

\begin{eqnarray}
\rho_1 & = & \frac{1}{4} ( \omega_1 + \omega_2 + \omega_3 + \omega_4
) \nonumber \\
\rho_2 & = & \frac{1}{4} ( - \omega_2  + \omega_4
) \nonumber \\
\rho_3 & = & \frac{1}{4} ( - \omega_1 + \omega_2 - \omega_3 +
\omega_4 ) \label{65}
\end{eqnarray}

while

\beq M^{AB} = \left( \begin{array}{cccc} \delta_1 & \delta_2 & \delta_3 & \delta_2 \\
\delta_2 & \delta_1 & \delta_2 & \delta_3 \\
\delta_3 & \delta_2 & \delta_1 & \delta_2 \\
\delta_2 & \delta_3 & \delta_2 & \delta_1 \end{array} \right)
\label{66} \eeq

with the following definitions

\begin{eqnarray}
\delta_1 & = & \frac{1}{2} \left(
\frac{\rho_1+\rho_3}{(\rho_1+\rho_3)^2 - 4 \rho^2_2} +
\frac{1}{\rho_1-\rho_3}
\right) \nonumber \\
\delta_2 & = &  - \frac{\rho_2}{(\rho_1+\rho_3)^2 - 4 \rho^2_2}
\nonumber \\
\delta_3 & = & \frac{1}{2} \left(
\frac{\rho_1+\rho_3}{(\rho_1+\rho_3)^2 - 4 \rho^2_2} -
\frac{1}{\rho_1-\rho_3} \right) \label{67}
\end{eqnarray}

Let us firstly discuss the ($1,3$) case.

The matrix $M_{AB}$ can be decomposed in terms of the following
submatrices

\beq M_{AB} = \left( \begin{array}{cc} M_{ab} & M_{a\alpha} \\
M_{\alpha b} & M_{\alpha\beta} \end{array} \right) = \left(
\begin{array}{c|ccc} \rho_1 & \rho_2 & \rho_3 & \rho_2 \\
\hline
\rho_2 & \rho_1 & \rho_2 & \rho_3 \\
\rho_3 & \rho_2 & \rho_1 &\rho_2 \\
\rho_2 & \rho_3 & \rho_2 & \rho_1 \end{array} \right) \label{68}
\eeq

while the matrix $M^{AB}$ is decomposed as

\beq M^{AB} = \left( \begin{array}{cc} M^{ab} & M^{a\alpha} \\
M^{\alpha b} & M^{\alpha\beta} \end{array} \right) =
\left( \begin{array}{c|ccc} \delta_1 & \delta_2 & \delta_3 & \delta_2 \\
\hline \delta_2 & \delta_1 & \delta_2 & \delta_3 \\
\delta_3 & \delta_2 & \delta_1 & \delta_2 \\
\delta_2 & \delta_3 & \delta_2 & \delta_1 \end{array} \right)
\label{69} \eeq

Moreover $\widetilde{M}^{ab}$ is the inverse of $M_{ab} = \rho_1$
and $\widetilde{M}_{ab}$ is the inverse of $M^{ab} = \delta_1$:

\beq \widetilde{M}^{ab} = \frac{1}{\rho_1} \ \ \ \ \ \
\widetilde{M}_{ab} = \frac{1}{\delta_1} \label{610} \eeq

Finally $\widetilde{M}^{\alpha\beta}$ is the inverse of
$M_{\alpha\beta}$

\beq \widetilde{M}^{\alpha\beta} = \frac{1}{D} \left(
\begin{array}{ccc} \rho_1^2 - \rho^2_2 & ( \rho_3 - \rho_1 ) \rho_2 & \rho^2_2 - \rho_1 \rho_3 \\
( \rho_3  - \rho_1 ) \rho_2 & \rho_1^2 - \rho_3^2 &
( \rho_3 - \rho_1 ) \rho_2 \\
\rho_2^2 - \rho_1 \rho_3 & ( \rho_3 - \rho_1 ) \rho_2 & \rho^2_1 -
\rho^2_2  \end{array} \right) \label{611} \eeq

with the determinant $D$ defined by

\beq D = \rho_1 ( \rho_1^2 - \rho_3^2 ) - 2 \rho^2_2 ( \rho_1 -
\rho_3 )  \label{612} \eeq

With these data we can compute the matrices

\begin{eqnarray}
& \ & K = \widetilde{M}_{ab} = \frac{1}{\delta_1} \nonumber \\
& \ & H = M_{a\alpha} \widetilde{M}^{\alpha\beta} M_{\beta b} =
\nonumber \\
& \ & = \frac{1}{D} [ \rho^2_3 ( \rho^2_1 - \rho^2_3 ) + 2 \rho^2_2
( \rho^2_1 + 2 \rho^2_3 - 3 \rho_1 \rho_3 ) ] \label{613}
\end{eqnarray}

Since $ D = \delta_1 ( \rho_1 - \rho_3 )^2 ( ( \rho_1 + \rho_3 )^2 -
4 \rho^2_2 ) $ we can deduce that

\begin{eqnarray} \lambda & = & \frac{\delta_1}{D} [ \rho^2_3 ( \rho^2_1 - \rho^2_3
) + 2 \rho^2_2 ( \rho^2_1 + 2 \rho^2_3 - 3 \rho_1 \rho_3 ) ] =
\nonumber \\ & \ & = \frac{(\rho_1+\rho_3) \rho^2_3 + 2 ( \rho_1 - 2
\rho_3 ) \rho^2_2 }{ ( \rho_1 - \rho_3 ) ( ( \rho_1 + \rho_3 )^2 - 4
\rho^2_2 ) } \label{614}
\end{eqnarray}

Let us notice here that in general in order to compute the
eigenvalues of the matrix $\Lambda^a_b$ we can substitute  $ M^{ac}
M_{c\beta} \widetilde{M}^{\beta\alpha} $ with $ - M^{a\alpha} $ and
we can finally obtain

\beq \Lambda^a_b = {(K^{-1})}^{ac} H_{cb} = - M^{a\alpha} M_{\alpha
b} = \delta_1 \rho_1 - 1 \label{615} \eeq

a formula already encountered in the case ($1,2$).

Let us discuss the case ($3,1$).

In this case the decomposition of the matrix $M_{AB}$ is different:

\beq M_{AB} = \left( \begin{array}{cc} M_{ab} & M_{a\alpha} \\
M_{\alpha b} & M_{\alpha\beta} \end{array} \right) = \left(
\begin{array}{ccc|c} \rho_1 & \rho_2 & \rho_3 & \rho_2 \\
\rho_2 & \rho_1 & \rho_2 & \rho_3 \\
\rho_3 & \rho_2 & \rho_1 &\rho_2 \\
\hline \rho_2 & \rho_3 & \rho_2 & \rho_1 \end{array} \right)
\label{616} \eeq

and analogously for the matrix $M^{AB}$

\beq M^{AB} = \left( \begin{array}{cc} M^{ab} & M^{a\alpha} \\
M^{\alpha b} & M^{\alpha\beta} \end{array} \right) =
\left( \begin{array}{ccc|c} \delta_1 & \delta_2 & \delta_3 & \delta_2 \\
\delta_2 & \delta_1 & \delta_2 & \delta_3 \\
\delta_3 & \delta_2 & \delta_1 & \delta_2 \\
\hline \delta_2 & \delta_3 & \delta_2 & \delta_1
\end{array} \right) \label{617} \eeq

The matrix $\widetilde{M}^{ab}$ is the inverse of $M_{ab}$

\beq \widetilde{M}^{ab} = \frac{1}{D(\rho_i)} \left(
\begin{array}{ccc} \rho^2_1 - \rho^2_2  & ( \rho_3 - \rho_1 ) \rho_2 &
\rho^2_2 - \rho_1 \rho_3 \\
( \rho_3  - \rho_1 ) \rho_2 & \rho^2_1 - \rho^2_3 &
( \rho_3 - \rho_1 ) \rho_2 \\
\rho^2_2 - \rho_1 \rho_3 & ( \rho_3 - \rho_1 ) \rho_2 & \rho^2_1 -
\rho^2_2  \end{array} \right) \label{618} \eeq

$\widetilde{M}_{ab}$ is the inverse of $M^{ab}$

\beq \widetilde{M}_{ab} = \frac{1}{D(\delta_i)} \left(
\begin{array}{ccc} \delta^2_1 - \delta^2_2  & ( \delta_3 - \delta_1 ) \delta_2 &
\delta^2_2 - \delta_1 \delta_3 \\
( \delta_3 - \delta_1 ) \delta_2 & \delta^2_1 - \delta^2_3 &
( \delta_3  - \delta_1 ) \delta_2 \\
\delta^2_2 - \delta_1 \delta_3 & ( \delta_3 - \delta_1 ) \delta_2 &
\delta^2_1 - \delta^2_2
\end{array} \right) \label{619} \eeq

$\widetilde{M}^{\alpha\beta}$ is the inverse of $M_{\alpha\beta} =
\rho_1$ e $\widetilde{M}_{\alpha\beta}$ is the inverse of
$M^{\alpha\beta} = \delta_1$:

\beq \widetilde{M}^{\alpha\beta} = \frac{1}{\rho_1} \ \ \ \ \ \ \
\widetilde{M}_{\alpha\beta} = \frac{1}{\delta_1} \label{620} \eeq

From these we can recover that $K_{ab} = \widetilde{M}_{ab}$ is the
inverse of $M^{ab}$ and therefore

\begin{eqnarray}
& \ & K^{-1}_{ab} = \left( \begin{array}{ccc} \delta_1 & \delta_2 &
\delta_3 \\ \delta_2 & \delta_1 & \delta_2 \\
\delta_3 & \delta_2 & \delta_1 \end{array} \right)
\nonumber \\
& \ & H_{ab} = M_{a\alpha} \widetilde{M}^{\alpha\beta} M_{\beta b} =
\frac{1}{\rho_1} \left( \begin{array}{c} \rho_2 \\
\rho_3 \\ \rho_2 \end{array} \right) ( \rho_2 \rho_3 \rho_2 )
\label{621}
\end{eqnarray}

Being a projector it has two null eigenvalues and only one $\neq 0
$.

\begin{eqnarray} \Lambda_{ab} & = & {(K^{-1})}^{ac} H_{cb} = \nonumber \\
& = & \frac{1}{\rho_1} \left( \begin{array}{c} \delta_1 \rho_2
+ \delta_2 \rho_3 + \delta_3 \rho_2  \\
2 \delta_2 \rho_2 + \delta_1 \rho_3 \\ \delta_1 \rho_2 + \delta_2
\rho_3 + \delta_3 \rho_2 \end{array} \right) ( \rho_2 \rho_3 \rho_2
) \nonumber \\
 & \ & \label{622} \end{eqnarray}

Let us notice that there is also the following simplification, due
to eq. (\ref{67})

\begin{eqnarray}
& \ & ( \delta_1 + \delta_3 ) \rho_2 + \delta_2 \rho_3
= - \delta_2 \rho_1 \nonumber \\
& \ & 2 \delta_2 \rho_2 + \delta_1 \rho_3  = - \delta_3 \rho_1
\label{623}
\end{eqnarray}

and therefore we obtain for the matrix $\Lambda_{ab}$

\beq \Lambda_{ab} = - \left( \begin{array}{c} \delta_2
\\ \delta_3 \\ \delta_2 \end{array} \right) ( \rho_2 \rho_3
\rho_2 ) \label{624} \eeq

Being of this form, all the matrix information is enclosed in its
trace

\beq Tr \Lambda_{ab} = - ( 2 \delta_2 \rho_2 + \delta_3 \rho_3 ) =
\delta_1 \rho_1 - 1 = \frac{(\rho_1+\rho_3) \rho^2_3 + 2 ( \rho_1 -
2 \rho_3 ) \rho^2_2 }{ ( \rho_1 - \rho_3 ) ( ( \rho_1 + \rho_3 )^2 -
4 \rho^2_2 ) } \label{625} \eeq

Being proportional to $ \rho^2_2 $ and $ \rho^2_3 $ the trace
becomes null when all the frequencies are equal between them.
$\Lambda_{ab}$ is a matrix with three eigenvalues, of which two null
and one equal to the case ($1,3$).

In general the cases ($1,N-1$) and ($N-1,1$) are solved by the
eigenvalue

\beq \lambda = \delta_1 \rho_1 - 1 \label{626} \eeq

where  $\rho_1 = \frac{1}{N}( \omega_1 + \omega_2 + .. + \omega_N )$
is the first entry of the matrix $M_{AB}$ and $ \delta_1 $ is the
first entry of the inverse matrix $M^{AB}$.

Finally we discuss the case ($2,2$).

In this case the matrix $M_{AB}$ is decomposed in the following
submatrices

\beq M_{AB} = \left( \begin{array}{cc} M_{ab} & M_{a\alpha} \\
M_{\alpha b} & M_{\alpha\beta} \end{array} \right) = \left(
\begin{array}{cc|cc} \rho_1 & \rho_2 & \rho_3 & \rho_2 \\
\rho_2 & \rho_1 & \rho_2 & \rho_3 \\ \hline
\rho_3 & \rho_2 & \rho_1 &\rho_2 \\
\rho_2 & \rho_3 & \rho_2 & \rho_1 \end{array} \right)
\label{627}\eeq

while the matrix $M^{AB}$ is decomposed as

\beq M^{AB} = \left( \begin{array}{cc} M^{ab} & M^{a\alpha} \\
M^{\alpha b} & M^{\alpha\beta} \end{array} \right) =
\left( \begin{array}{cc|cc} \delta_1 & \delta_2 & \delta_3 & \delta_2 \\
\delta_2 & \delta_1 & \delta_2 & \delta_3 \\ \hline
\delta_3 & \delta_2 & \delta_1 & \delta_2 \\
\delta_2 & \delta_3 & \delta_2 & \delta_1 \end{array} \right)
\label{628} \eeq

The entropy calculation is in this case solved if we know the
eigenvalues of the following matrix

\beq \Lambda^a_b  = - M^{a\alpha} M_{\alpha b} = - \left( \begin{array}{cc} \delta_3 & \delta_2 \\
\delta_2 & \delta_3 \end{array} \right)
\left( \begin{array}{cc} \rho_3 & \rho_2 \\
\rho_2 & \rho_3 \end{array} \right) \label{629} \eeq

The main reason why the eigenvalues of such a matrix $\Lambda^a_b$
are real is because it can be rewritten as the product of two
symmetric matrices of which at least one with all positive
eigenvalues. This property assures that the two matrices can be
simultaneously diagonalized.

Let us notice that the matrix $M_{\alpha b}$ can be diagonalized
with the transformation

\beq \left( \begin{array}{cc} \rho_3 & \rho_2 \\
\rho_2 & \rho_3 \end{array} \right)  =
\left( \begin{array}{cc} a & - b \\
b & a \end{array} \right)
\left( \begin{array}{cc} \rho_3 + \rho_2   & 0 \\
0 & \rho_3 - \rho_2  \end{array} \right)
\left( \begin{array}{cc} a & b \\
- b & a \end{array} \right) \label{630} \eeq

where $ a = b = \frac{1}{\sqrt{2}} $, from which

\beq M^{a\alpha} M_{\alpha b} = \left( \begin{array}{cc} a & - b \\
b & a \end{array} \right) \left( \begin{array}{cc} \delta_3+\delta_2 & 0 \\
0 & \delta_3-\delta_2 \end{array} \right)
\left( \begin{array}{cc} \rho_3 + \rho_2 & 0 \\
0 & \rho_3 - \rho_2 \end{array} \right)
\left( \begin{array}{cc} a & b \\
- b & a \end{array} \right) \label{631} \eeq

In this particular case the two matrices are simultaneously
diagonalized by the same invertible matrix and we can directly
compute the eigenvalues

\beq \lambda_{\pm} = - ( \delta_3 \pm \delta_2 ) ( \rho_3 \pm \rho_2
) = \frac{ ( \delta_1 \pm \delta_2 )}{ ( \rho_1 \pm \rho_2 )} (
\rho_3 \pm \rho_2 )^2 \label{632} \eeq

Therefore $\lambda_{\pm}$ are real eigenvalues and the resulting
entropy is real and positive.

\section{Continuum limit}

Until now we have limited ourself to the study of the entanglement
of a scalar field reduced to a lattice of $N$ point where the total
entropy of the system is always finite. The continuum limit can be
worked out by sending the volume to infinity and keeping the lattice
spacing $a$ finite, as in \cite{17}. In that case the entropy
receives contributions from an integral operator admitting a
spectrum of infinite eigenvalues accumulating around zero. The exact
calculation goes beyond our possibilities and we are obliged to
confirm the observations made in \cite{17}-\cite{18}, i.e. that it
is possible to single out a finite surface entropy ( proportional to
the boundary area ). In particular in \cite{17}-\cite{18} it is
shown how to factor out in four dimensions an entropy per unit area:

\beq S/A  = C a^{-2} \eeq

where $C$ is some constant and $a$ is the momentum cutoff or lattice
spacing. The proportionality constant $C$ has the right order of
magnitude if $a$ is of the order of the Planck length $l_p$.
Although in general the system is not exactly solvable, it is
possible to make some numerical estimate on the upper bound of this
surface entropy.

\section{Discussion}

The entanglement method has been revealed effective to confirm the
results obtained with the brick wall method. In this work we have
firstly noticed a greater resemblance between the two methods
clarifying that the inaccessible quantum degrees of freedom
contributing to black hole entropy are situated inside the brick
wall and not in the interior of black hole. We have pointed out that
in two dimensions there is a coordinate transformation flattening
the background metric and in the massless limit the quantization of
the scalar field in the gravitational background can be reduced to
the free field case.

Then we have quantized the system reduced to a finite number $N$ of
degrees of freedom and computed in several examples their
contribution to entanglement entropy finding some general result
that may be useful in the large $N$ limit. In these cases our model
is an explicit application of the general method outlined in
\cite{17}.

The ambitious goal would be to solve exactly the spectrum of the
integral operator generalizing our results in the large $N$ limit
and to check if there is a contribution to entropy proportional to
the area of the spatial division, but the technical difficulties are
such that we are obliged to conclude our study here, postponing the
large $N$ exact solution to a future research.

\end{document}